\documentclass[useAMS,usenatbib]{mn2e}
\usepackage{graphicx}
\usepackage{amssymb}

\newcommand{\placefigone}{
\begin{figure}
\begin{center}
\includegraphics[width=\columnwidth]{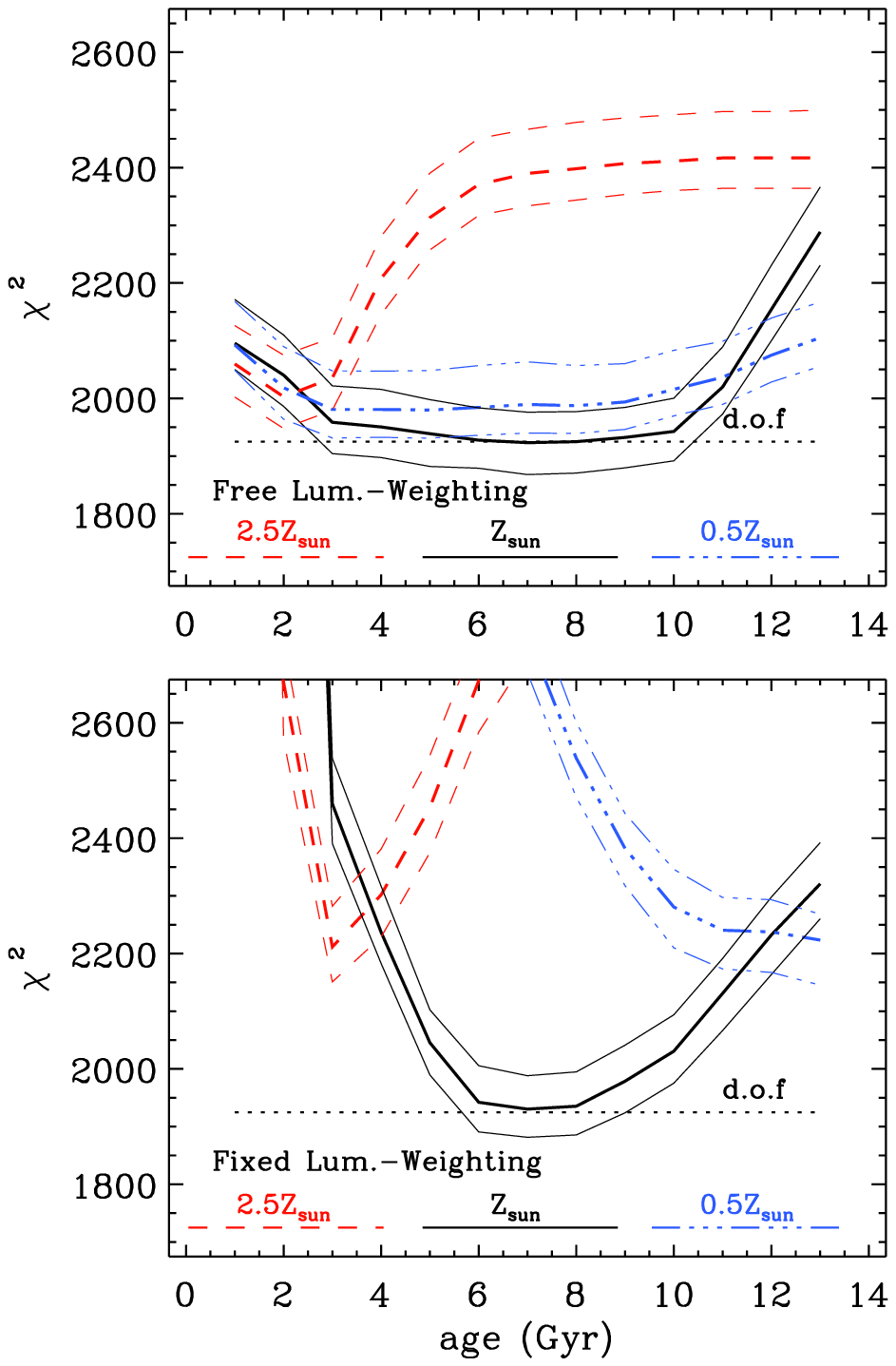}
\end{center}
\caption{
Simulations showing the accuracy in recovering the age of a 7-Gyr-old
disc population that is embedded in a 13-Gyr-old bulge population. The
bulge and disc populations contribute with the same broad-band flux
between 4500~\AA\ and 5500~\AA\ to the input model, and are represented
by single-age stellar population models of solar metallicity from
\citet{Mar11}, covering the entire wavelength range of our VIMOS
observations (\S~\ref{sec:analysis}).
Different amounts of kinematical broadening were also included in the
bulge and disc models ($\sigma$ = 200 and 100$\rm km\;s^{-1}$), and
statistical fluctuations were added to simulate a $S/N\sim100$. In
both panels the thick solid line shows the average quality of the fit
to the input model as disc single-age models of increasing ages are
combined with the right, 13-Gyr-old, bulge population model, which
mimics the situation where the bulge stellar properties have been
previously constrained from off-centred observations. The thin solid
lines indicate the 1$\sigma$ fluctuation that is derived by fitting
different realisations of the input model. The dashed and dot-dashed
lines show the quality of the fits obtained while considering disc
populations of super- and sub-solar metallicities.
On the left panel the relative contribution to the fit of the bulge
and disc populations is left free to vary, whereas on the right panel
each component is constrained to contribute the same broad-band flux
in the 4500~\AA--5500~\AA\ range as in the input model. In both cases
the kinematic broadening of the templates is also freely adjusted
during the fit.
The comparison between $\chi^2$ curves in the left and right panels
illustrates how knowing in advance the disc-light contribution allows
one to recover the correct disc age and metallicity, whereas without
such an initial clue estimating the disc age and metallicity would be
complicated by the well-known degeneracy between these two parameters.
}
\label{fig:sim}
\end{figure}
}

\newcommand{\placefigtwo}{
\begin{figure}
\begin{center}
\includegraphics[width=\columnwidth]{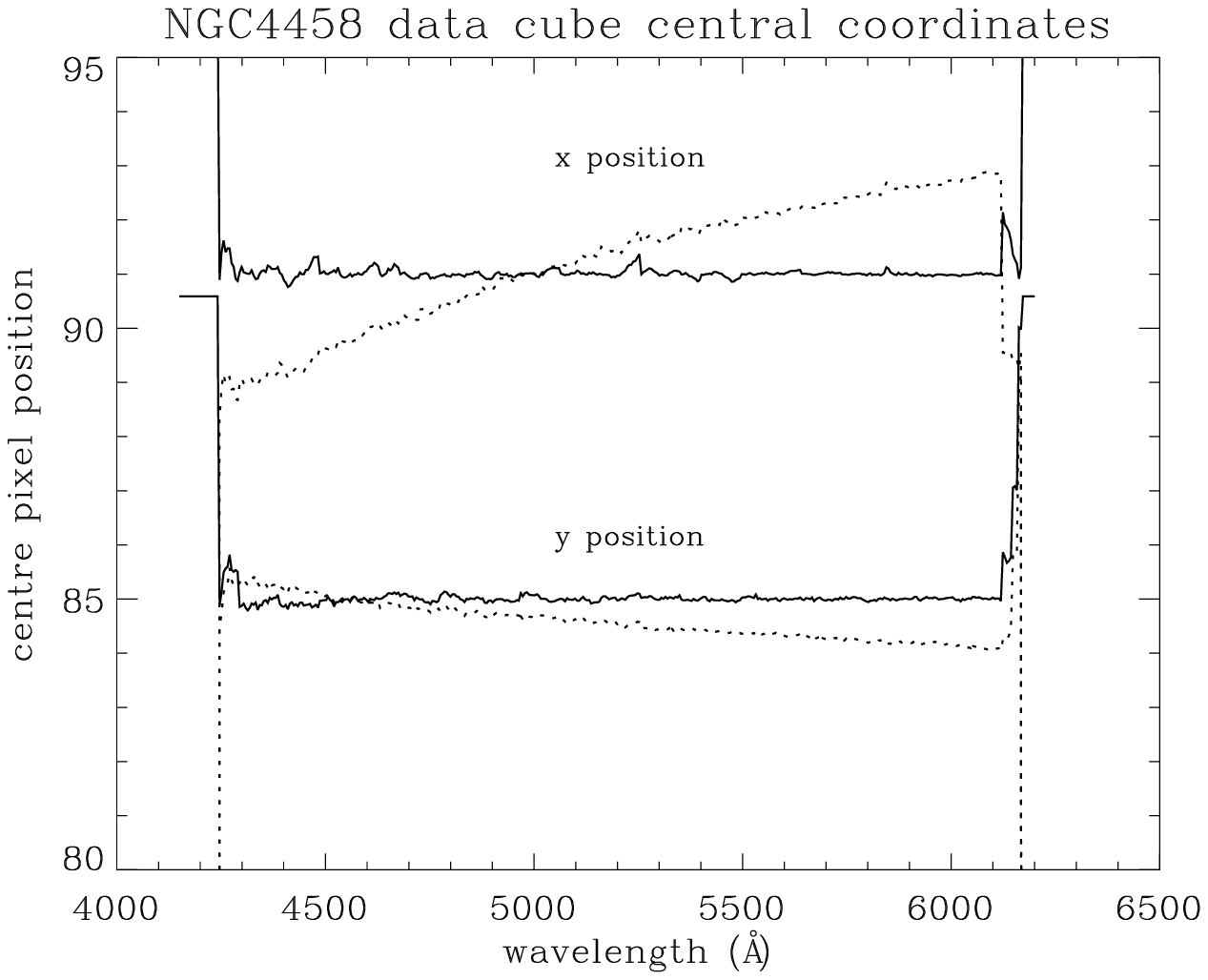}
\end{center}
\caption{The $x$ and $y$ centre coordinates of NGC~4458 as a function
  of wavelength along the VIMOS data cube before (dotted lines) and
  after our extra rectification (solid lines).}
\label{fig:rectif}
\end{figure}
}

\newcommand{\placefigthree}{
\begin{figure}
\begin{center}
\includegraphics[width=\columnwidth]{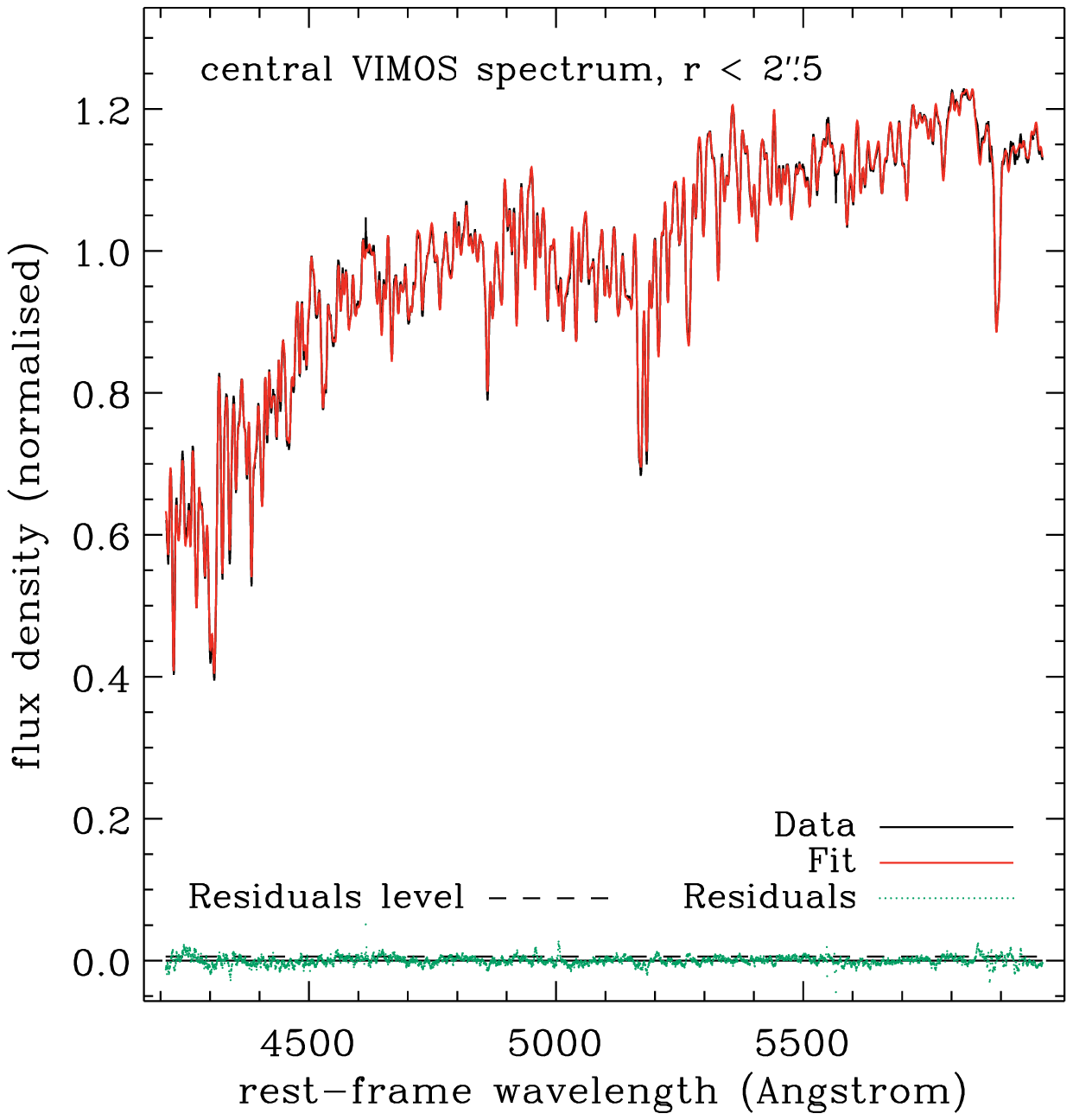}
\end{center}
\caption{Central 5\arcsec\ VIMOS spectrum of NGC~4458 and our best
  pPXF fit using the entire MILES spectral library. The level of
  fluctuations in the residuals of this fit correspond to an average
  value of 175 for the signal over residual-noise ratio.}
\label{fig:censpec}
\end{figure}
}

\newcommand{\placefigfour}{
\begin{figure}
\begin{center}
\includegraphics[width=\columnwidth]{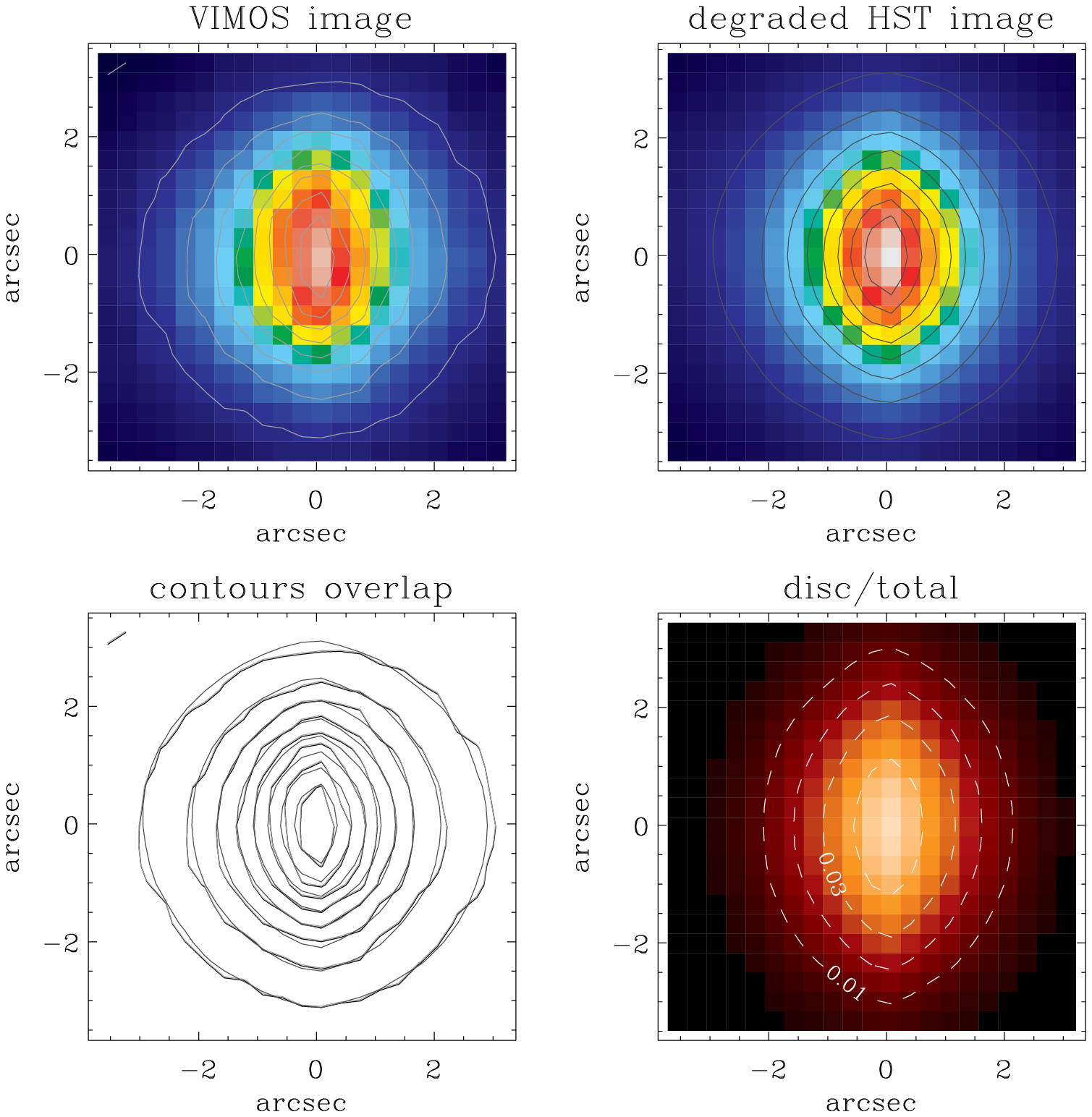}
\end{center}
\caption{Upper panels: Central parts of the VIMOS reconstructed image
  of NGC~4458 (left) and the same for the F555W WFPC2 image, once
  degraded to match the spatial resolution of VIMOS and resampled
  within the 0\farcs33$\times0$\farcs33 resolution elements of the
  high-resolution configuration of VIMOS (right). Lower left: overlap
  of the surface brightness contours for the VIMOS and degraded HST
  image (in light and dark grey lines, respectively), which are also
  shown in the corresponding upper panels. Lower right: map for the
  disc-to-total light ratio computed from the HST degraded image shown
  in the upper right panel and a similarly convolved and resampled
  version for the image of the best-fitting disc model that was
  obtained during the Scorza \& Bender disc-bulge decomposition of
  \citet{Mor04}. The dashed contours on this last image show the level
  of disc-light contribution in the VIMOS spaxels (but see text).
  Note that the major axis of NGC~4458 and its nuclear disc runs
  nearly parallel to the x-axis of the VIMOS cube, and the vertical
  elongation shown here is the results of the rather elongated
  point-spread function of VIMOS and the steep profile of the
  intrinsic surface brightness of this galaxy.}
\label{fig:hst_vimos}
\end{figure}
}

\newcommand{\placefigfive}{
\begin{figure}
\begin{center}
\includegraphics[width=\columnwidth]{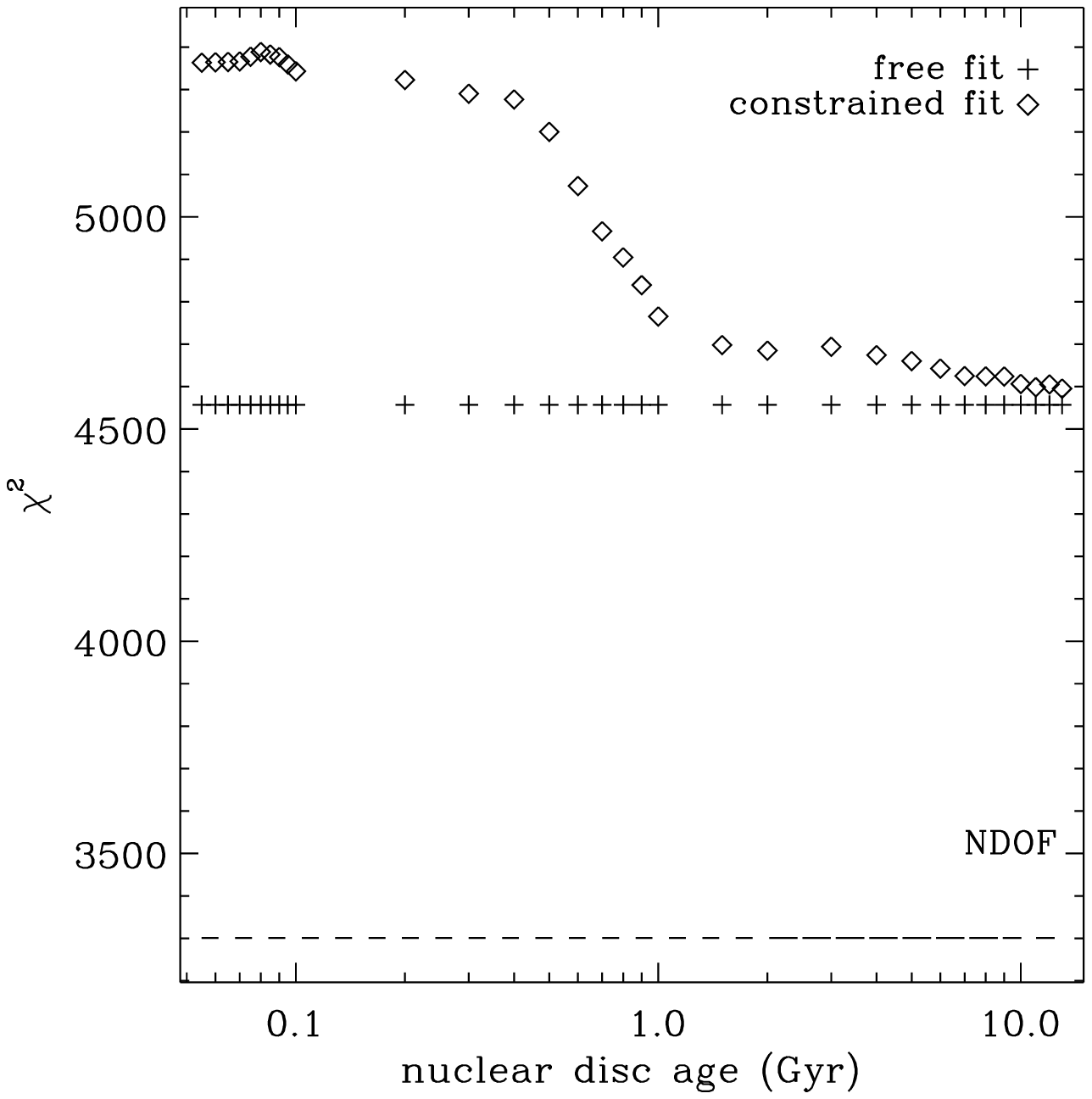}
\end{center}
\caption{Quality of our constrained (diamonds) and free (plus signs)
  fits as a function of the stellar age of the NSD, or more
  specifically, of the single-age stellar population model for disc
  that is combined with our empirical bulge template.  Even though the
  quality of these fits is relatively poor compared to the quality of
  the fit that can be achieved through a free combination of all the
  stellar templates in the MILES library (as done for the bulge
  spectrum in \S~\ref{subsec:analysisBulge}, horizontal dashed line),
  the observed trends for our constrained indicates a very old stellar
  age for the nuclear disc. In fact, this is also the case of the free
  fit, since only the bulge template is used. All $\chi^2$ values were
  computed after rescaling the formal uncertainties on the flux
  densities of our nuclear spectrum so that the best fit based on the
  entire MILES stellar library (\S~\ref{subsec:analysisNSD}) sets our
  standard for a good fit, thus making the corresponding
  $\chi^2$=NDOF.}
\label{fig:first_results}
\end{figure}
}

\newcommand{\placefigsix}{
\begin{figure}
\begin{center}
\includegraphics[width=3.3in]{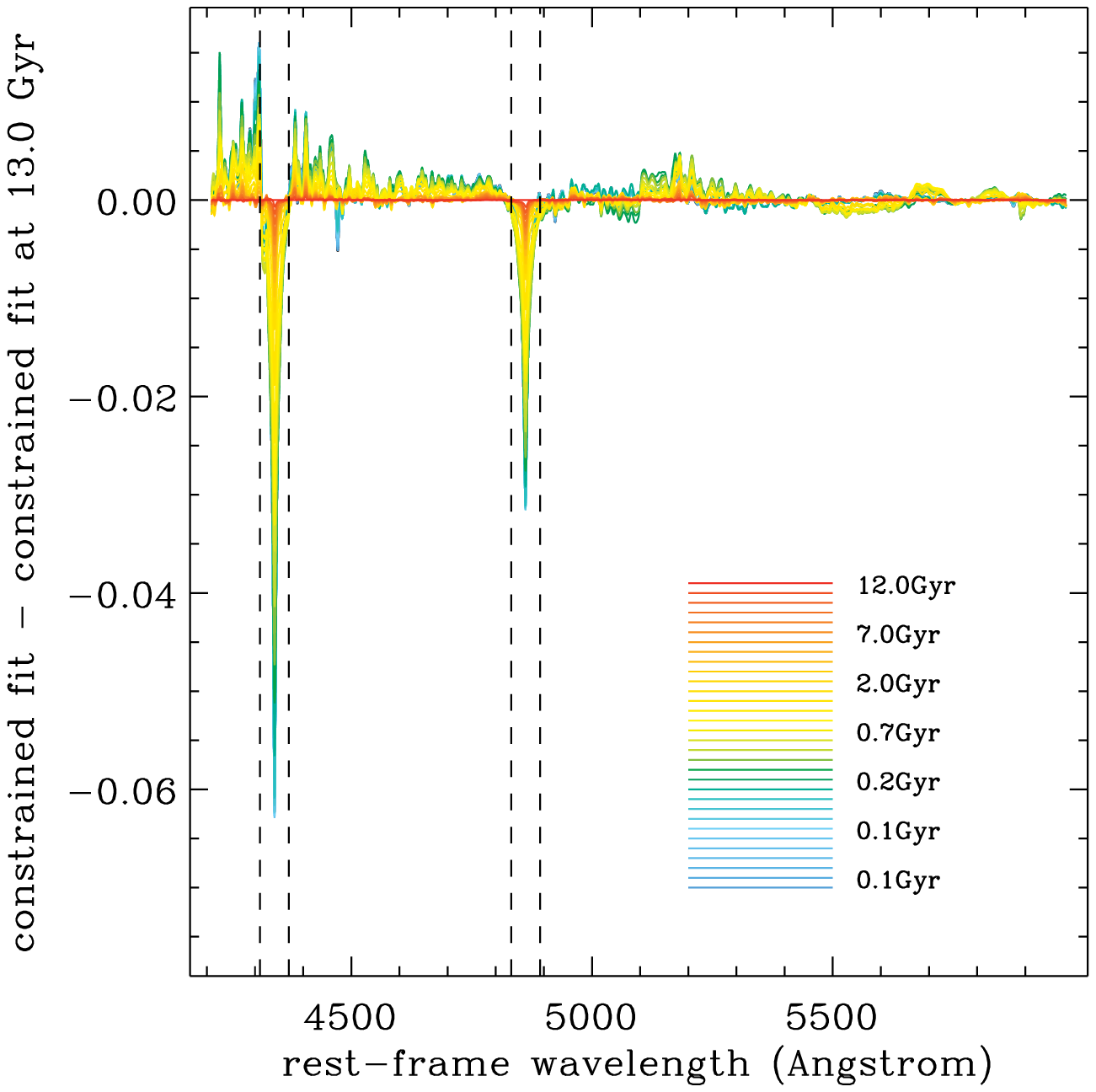}
\end{center}
\caption{Difference between the best-fitting constrained fit model,
  which features a 12.6-Gyr-old disc, and all other constrained fits
  including different single-age disc templates. The largest
  deviations are observed within the spectral regions identified by
  the vertical dashed lines, which correspond to the age-sensitive
  H$\delta$ and H$\beta$ stellar absorption features (at 4340~\AA\ and
  4861~\AA, respectively). Nearly 90\% of the quadratic
  difference (relevant for comparing model $\chi^2$ values) with the
  youngest disc model is contained across these 30~\AA-wide spectral
  windows, with this fraction lowering to $\sim$40\% when comparing
  our best model with other models featuring old discs.}
\label{fig:model_diffs}
\end{figure}
}

\newcommand{\placefigseven}{
\begin{figure}
\begin{center}
\includegraphics[width=3.3in]{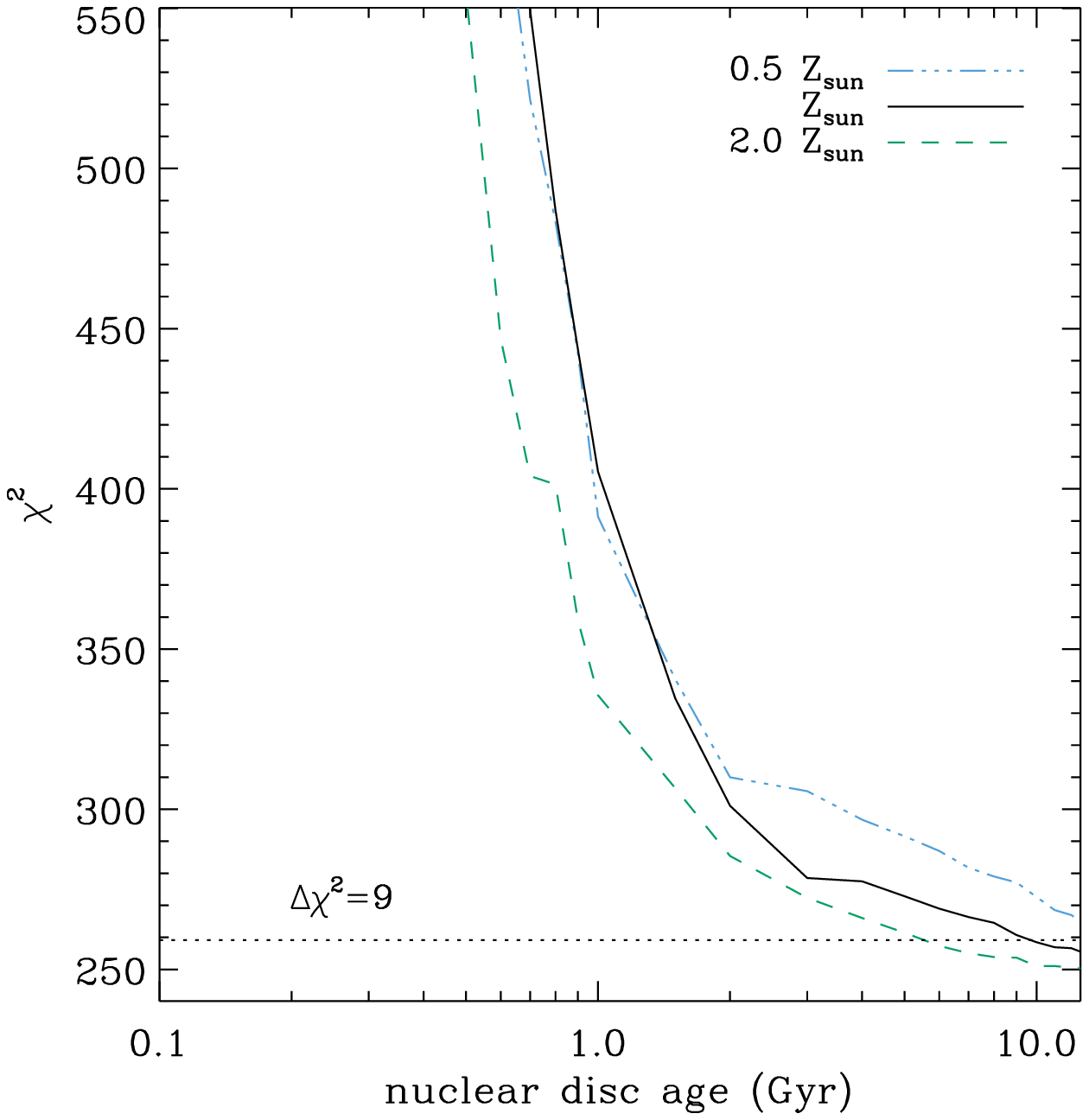}
\end{center}
\caption{Final stellar age estimate for the NSD of NGC~4458. Similar
  to Fig.~\ref{fig:first_results}, but now based only on constrained
  fit model featuring single-age disc templates of different
  metallicities (solid lines for Solar values, dot-dashed and dashed
  for half and twice Solar values, respectively) and while assessing
  the quality of the fits within the H$\delta$ and H$\beta$ spectral
  windows shown in Fig.~\ref{fig:model_diffs}.
  All plotted $\chi^2$ values within this spectral region have been
  computed after rescaling all flux density errors assuming that the
  best constrained fit model would yield a $\chi^2$=NDOF across the
  {\it entire\/} spectrum.
  A $\Delta\chi^2=9$ threshold set from the best $\chi^2$ value
  obtained within the H$\delta$ and H$\beta$ windows indicate that at
  best, when considering populations of super-Solar metallicity, the
  nuclear disc could have formed as recently as $\sim$5--6 Gyr ago.}
\label{fig:final_results}
\end{figure}
}

\newcommand{\placefigeight}{
\begin{figure}
\begin{center}
\includegraphics[width=3.3in]{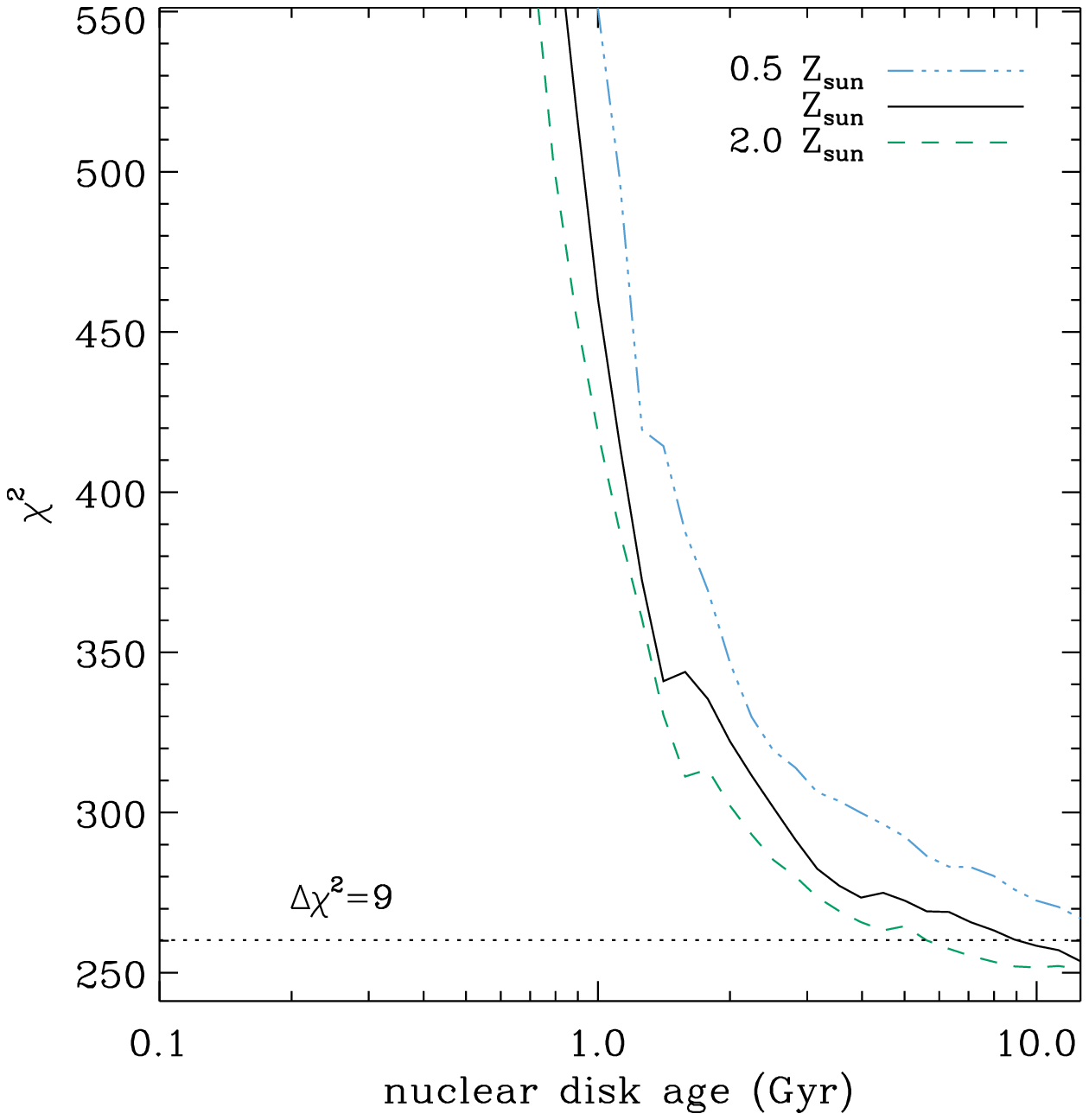}
\end{center}
\caption{Similar to Fig.~\ref{fig:final_results} but now while
  adopting the single-age models of \citet{Vaz10}, based also on the
  MILES stellar library.}
\label{fig:final_results_M11}
\end{figure}
}

\title[The Nuclear Disc of NGC~4458]{Nuclear discs as clocks for the
  assembly history of early-type galaxies: the case of NGC~4458}

\author[M. Sarzi et al.]{\parbox{\textwidth}{M. Sarzi$^{1}$\thanks{E-mail:
    m.sarzi@herts.ac.uk}, H.~R. Ledo$^{1}$, L. Coccato$^{2}$,
  E.~M. Corsini$^{3,4}$, M. Dotti$^{5,6}$, S. Khochfar$^{7}$,
  C. Maraston$^{8}$, L. Morelli$^{3,4}$, A. Pizzella$^{3,4}$}\vspace{0.4cm}\\
\parbox{\textwidth}{
$^{1}$Centre for Astrophysics Research, University of Hertfordshire,
  College Lane, Hatfield AL10 9AB, United Kingdom\\
$^{2}$ESO, Karl-Schwarzschild-Strasse 2, 85748, Garching, Germany\\
$^{3}$Dipartimento di Fisica e Astronomia ``G. Galilei'', Universit\`a di
Padova, vicolo dell'Osservatorio 3, I-35122 Padova, Italy \\
$^{4}$INAF-Osservatorio Astronomico di Padova, vicolo dell'Osservatorio 5,
I-35122 Padova, Italy\\
$^{5}$Dipartimento di Fisica G. Occhialini, Universit\`a degli Studi
di Milano, Bicocca, Piazza della Scienza 3, 20126 Milano, Italy\\
$^{6}$INFN, Sezione di Milano-Bicocca, Piazza della Scienza 3, 20126
Milano, Italy\\
$^{7}$Institute for Astronomy, University of Edinburgh, Royal Observatory,
Blackford Hill, Edinburgh EH9 3HJ, United Kingdom\\
$^{8}$Institute of Cosmology and Gravitation, University of Portsmouth,
Dennis Sciama Building, Portsmouth PO1 3FX, United Kingdom }
}

\begin{document}


\pagerange{\pageref{firstpage}--\pageref{lastpage}} \pubyear{2013}

\maketitle

\label{firstpage}

\begin{abstract}
Approximately 20\% of early-type galaxies host small nuclear stellar
discs that are tens to a few hundred parsecs in size. Such discs
are expected to be easily disrupted during major galactic encounters,
hence their age serve to constrain their assembly
history.
We use VIMOS integral-field spectroscopic observations for the {\it
  intermediate-mass\/} E0 galaxy NGC~4458 and age-date its nuclear
disc via high-resolution fitting of various model spectra. We find
that the nuclear disc is at least 6 Gyr old. A clue to gain narrow
limits to the stellar age is our knowledge of the nuclear disc
contribution to the central surface brightness.
The presence of an old nuclear disk, or the absence of disruptive
encounters since $z\sim0.6$, for a small galaxy such as NGC~4458 which
belongs to the Virgo cluster, may be consistent with a hierarchical
picture for galaxy formation where the smallest galaxies assembles
earlier and the crowded galactic environments reduce the incidence of
galaxy mergers. On the other hand, NGC~4458 displays little or no bulk
rotation except for a central kpc-scale kinematically-decoupled
core. Slow rotation and decoupled core are usually explained in terms
of mergers. The presence and age of the nuclear disc constraint these
mergers to have happened at high redshift.
\end{abstract}

\begin{keywords}
galaxies: elliptical and lenticular, cD -- galaxies: evolution,
galaxies: nuclei, galaxies : structure, galaxies: formation
\end{keywords}

\section[]{Introduction}
\label{sec:intro}

In the context of a dark-matter dominated Universe galaxies should
have grown through a combination of star formation and merging
processes, whereby on the one hand star formation was regulated by the
presence of fresh gaseous material and the negative feedback of
supernovae and possibly also active nuclei, and on the other hand
merging events should have proceeded in a hierarchical fashion.
A steady increase in the quality and scope of spectroscopic
investigations of nearby galaxies has allowed to constrain directly
their star-formation history \citep[e.g.,][for early-type
  galaxies]{Tho05,Tho10}, but reconstructing the assembly history of
galaxies has proved so far to be more difficult.
Attempts to quantify the rate of merging events by searching close
galactic pairs and interacting galaxies \citep[e.g.,][]{Dar10} depend
on the depth and the area covered by the images used in these studies,
whereas individual investigations of morphological signatures of past
mergers, such as galaxy shells, can hardly pinpoint the epoch of such
a galactic encounter \citep[but see][]{Hau99}.
The lack of constraints on the assembly history of nearby galaxies
leaves unchecked several predictions of the hierarchical standard
paradigm. For instance, the most massive galaxies should have
assembled only recently \citep[e.g.,][]{DeL06,Kho06} whereas, at a
given mass, galaxies in clusters should have experienced less merging
events than their counterparts in the field since when they entered
such crowded galactic environments. Indeed in clusters galaxies fly
too fast by each other to merge efficiently.

Understanding the assembly history of galaxies is particularly
important in the case of early-type galaxies, as these systems have
long been thought to originate during merging events \citep{Tom77}.
In fact, the kinematic distinction between fast and slowly-rotating
early-type galaxies, first suggested with long-slit data
\citep[e.g.,][]{Ill77,Bin78,Dav83,Ben94} and recently quantified
thanks to integral-field data \citep{Ems07,Cap07} is suggestive also
of a separate merging history for these two kinds of objects
\citep{Ems11,Kho11,Naa14}, where slow rotators would owe their low
angular momentum to a more systematic bombardment by smaller satellite
galaxies.

In this respect, nuclear stellar discs (NSDs) could prove important
tools to directly constrain the assembly history of early-type
galaxies. Initially discovered in images taken with the Hubble Space
Telescope \citep[HST; ][]{vdB94} and now known to be common in
early-type galaxies \citep[in up to 20\% of them][]{Led10}, such small
discs (a few $\sim 100\rm pc$ in radius at most) are indeed fragile
structures that should not survive a significant merger event (see,
e.g., the simulations shown in Sarzi, Ledo \& Dotti 2015). This means
that by dating the stellar age of the NSDs it is possible to place a
lower limit for the look-back time since their host galaxies
experienced a major encounter, as NSDs could form also after such an
event.
In fact, the stellar age of NSDs can be constrained even more
precisely than is generally the case for other kinds of galactic
component, thanks to the possibility to derive in advance their
relative contribution to the total galaxy light.

The main difficulty in disentangling a superposition of two stellar
populations in the spectra of a galaxy, in this case the nuclear disc
and the bulge, is the degeneracy between the age and light fractions
of each component. Using good-quality and extended spectra allows to
better exploit the information encoded in the stellar absorption lines
and can help mitigating this problem, but in the presence of
relatively old stellar populations further complications arise from
the degeneracy between age and metallicity or reddening.
On the other hand, in the case of structurally different and, to some
extent, well described, galactic components it may be possible to
infer from images their individual light fractions in the considered
spectra (in the wavelength range covered by the images), and exploit
this constrain to break the previous degeneracies.
This is precisely the case of a nuclear disc embedded in a stellar
bulge, where the surface brightness distribution of the disc can be
inferred using the disc-bulge decomposition technique introduced by
\citet{Sco95}, which relies only on the assumption of an exponential
radial profile for the disc and an elliptical shape for the bulge
isophotes \citep[see also, e.g.,][]{Piz02,Mor04,Mor10,Cor12}.

\placefigone

Simple simulations such as those shown in Fig.~\ref{fig:sim} serve to
illustrate the dramatic effect that an {\it a priori\/} knowledge of
the disc light contribution should have in estimating the age of NSD
embedded in a bulge. Fig.~\ref{fig:sim} shows the case of a 7-Gyr-old
disc population embedded in a 13-Gyr-old bulge, where both components
contribute to the input model spectrum with the same broad-band flux
in the 4500~\AA\ -- 5500~\AA\ range and are represented by single-age
stellar population models of solar metallicity \citep[from][]{Mar11}.
When trying to match different noisy realisations of such input model
by combining the correct bulge template (as if the bulge stellar
properties had also been previously constrained) with disc model
populations of varying age and metallicity, the $\chi^2$ contours
around the input disc stellar age of 7 Gyr increase steeply to high
values if the relative contribution of the bulge and disc component
are fixed to their input value (as if they were known from a
disc-bulge decomposition, Fig.~\ref{fig:sim} right panel). On the
other hand, when the relative contribution of the disc and bulge
templates are not constrained it is possible to obtain a very good
match to the input spectrum also when using disc populations of
considerably different age and metallicity than the input that of the
input disc template (Fig.~\ref{fig:sim}, left panel).
In other words, this experiment shows that knowing in advance the disc
light contribution should allow estimating more robustly the NSD age
and metallicity.

Motivated by the potential use of NSDs as clocks for the assembly
history of their host galaxies and encouraged by the previous kind of
simulations, this paper presents a pilot investigation based on
integral-field spectroscopic observations of the NSD in the Virgo
elliptical galaxy NGC~4458, which has a well-known nuclear disk
\citep[][]{Mor04,Mor10} and thus constitutes an ideal laboratory for
testing how accurately we can estimate the age of NSDs.
This paper is organised as follows. In \S~\ref{sec:obsdatared} we
describe our observations with the Very Large Telescope (VLT) and the
reduction of our data taken with the VIsible MultiObject Spectrograph
(VIMOS). The core of our analysis is found in \S~\ref{sec:analysis},
where start by estimating the relative contribution of the bulge and
NSD light in our central VIMOS spectra (\S\ref{subsec:analysisHST}),
extract a template for stellar bulge population
(\S~\ref{subsec:analysisBulge}) and finally constrain the stellar age
of the nuclear disc (\S~\ref{subsec:analysisNSD}). Finally, in
\S~\ref{sec:discuss} we discuss our results, suggesting also some
future avenues for the methodology developed here
(\S~\ref{subsec:discussfuture}).

\section{VIMOS observations and data reduction}
\label{sec:obsdatared}

\subsection{Observations}
\label{subsec:obs}

The VIMOS Integral Field Unit \citep{LeF03}, installed on the
Melipal-UT3 of the VLT, presented itself as one of the best
instruments for this study.
With VIMOS it is indeed possible to extract spectra of intermediate
spectral resolution over a relatively long wavelength range and at
different spatial locations, which allows to study the stellar
population of NSDs while constraining also the properties of the
surrounding stellar bulge. Yet, it is the large collecting power of
VLT and the possibility to obtain observations under the best seeing
conditions that make VIMOS ideal to study NSDs, as these contribute
significantly to the total stellar light only over very small spatial
scales.

The VIMOS data for NGC~4458 were collected in service mode on April
2007 (P79) and between April and June 2008 (P81), using the HR blue
grating with no filter and while opting for highest spatial
magnification. Such a configuration lead to datacubes comprising of
1600 spectra extending from 4150~\AA\ to 6200~\AA\ and with spectral
resolution of 2~\AA\ (FWHM), each sampling an area
$0\farcs33\times0\farcs33$ within a total field-of-view of
$13\farcs0\times13\farcs0$.
To allow for a proper sky subtraction and minimise the impact the dead
fibres or pixels, each observing block consisted of two, slightly
offset on-source pointing (each 940s long in P79 and 1025s in P81),
bracketing a shorter sky exposure (for 480s and 500s in P79 and P81,
respectively).
Out of a total allocated time of 23h, considering that only one
observing block was executed in P79, this strategy yielded a total of
5.7h on target.
All these observations were taken under very good atmospheric seeing
conditions, on average around $0\farcs8$, and at an average airmass of
$\sim 1.33$.

\subsection{Data reduction}
\label{subsec:datred}

We started the reduction of our data by running each of our single sky
and on-target exposures through the VIMOS ESO
pipeline\footnote{version 2.2.1 {\tt
    http://www.eso.org/sci/software/pipelines/\/}}, thus carrying out
the bias subtraction, flat fielding, fibre identification and tracing,
and wavelength calibration.
We then used in-house IDL and IRAF procedures to further correct for
the different relative transmission of the VIMOS quadrants, which we
adjusted by requiring the same intensity for the night-sky lines
across the field-of-view, and in order to subtract the sky spectrum
from the galaxy pointings. During this last step, we compensate for
time variations in the night sky spectrum between the on-target and
sky pointing by adjusting the strength of the strongest night sky
lines in the sky exposures to match what found in the galaxy
pointings.
Finally, each on-target exposure was organised in data cubes using the
tabulated position in the field-of-view of each fibre, which were then
merged in a final data cube by aligning the bright nuclear regions of
NGC~4458 in the total reconstructed images corresponding to each
single cube.

\placefigtwo

Although these steps should have sufficed in providing a fully reduced
data cube, we noticed that an extra rectification was needed to
account for a residual systematic shift of the galaxy centre as we
move along the wavelength direction in the data cube.  Such a shift is
likely due to atmospheric differential refraction, as the galaxy
centre moves mostly along the $x$ axis of our data cube and that this
is close to north-south direction.  Fortunately, the nuclear regions
of NGC~4458 are sufficiently cuspy for us to accurately locate the
galaxy centre as a function of wavelength, thus correcting for this
systematic shift. Fig.~\ref{fig:rectif} shows the galaxy centre $x$
and $y$ pixel coordinates as a function of wavelength, before and
after rectifying our final data cube. Even though the centre of
NGC~4458 moved only by $1\farcs5$ between the blue and red ends of our
data cube, correcting for this shift is particularly important in the
context of this work, since the NSD of this galaxy contributes
significantly to the central light distribution only within a few tens
of an arcsecond (see \citealp{Mor04}). In fact, ensuring an accurate
rectification meant restricting the final wavelength range of our data
cube between 4220~\AA\ and 6000~\AA.
The final quality of our data can be appreciated in
Fig.~\ref{fig:censpec}, where we show the integrated spectrum of
NGC~4458 within the central 5\arcsec.
Based on the residuals of the best-fitting model for the stellar
spectrum, obtained the pixel-fitting code of \citep[][pPXF]{Cap04} and
the whole MILES stellar library \citep{San06}, we obtain an average
value of 175 for the signal over residual-noise ratio.
For comparison, within a similar aperture and for 2h of on-source
exposure time, the SAURON integral-field data for NGC~4458, first
presented by \citep{Ems04}, come with a signal over residual-noise
ratio $\sim$250 in the 4850 -- 5300~\AA\ wavelength region.

\placefigthree
\placefigfour

\section{Analysis}
\label{sec:analysis}

As we covered in the introduction, in order to best estimate the
stellar age of the nuclear disc in NGC~4458 we ought to know the
relative contribution of the disc and its surrounding bulge to the
central stellar surface brightness distribution, preferably within a
wavelength range that is covered by our spectra.
For this we will use the results of the disc-bulge decomposition of
\citet{Mor04}, which was based on HST-F555W images in the visible
domain, and by accounting for the difference in spatial resolution and
spatial sampling between HST and VIMOS we will estimate the disc
contribution to each of our central VIMOS spectra
(\S~\ref{subsec:analysisHST}).
Thanks to the integral-field nature of our data, we will then extract
a central aperture spectrum while striking a good compromise between
signal-to-noise ratio and disc light contribution, as well as an
off-centered aperture spectrum dominated by the bulge light that could
be used as a template for such a component
(\S\ref{subsec:analysisBulge}) in our final stellar-population
analysis of the nuclear regions (\S\ref{subsec:analysisNSD}).

\subsection{HST to VIMOS Matching}
\label{subsec:analysisHST}

Starting from the disc-bulge decomposition of \citeauthor{Mor04}, we
can account for the lower spatial (seeing-limited) resolution and the
coarser spatial sampling of the VIMOS observations to compute the
fraction of disc light that would have been observed within each VIMOS
resolution element, if using the same filter of the HST images.
We started such a matching procedure by rotating the HST image of
NGC~4458 to match the orientation of the VIMOS reconstructed
image. Then, after extracting only its central regions, we proceeded
to convolve the HST image by a double-Gaussian meant to represent the
atmospheric point-spread function (PSF) of VIMOS and finally resampled
the resulting degraded image within the 0\farcs33$\times0$\farcs33
VIMOS spaxels.
To match the VIMOS reconstructed image this procedure required a
rather vertically elongated PSF, which effectively greatly reduces our
spatial resolution in that direction. This problem is only briefly
mentioned in the VIMOS documentation and would appear to be due to the
placing of the IFU unit at the edge of the VIMOS field of view
\citep{Ang08}. On the other hand, the extent of the PSF along the
horizontal direction would appear to remain at the nominal level we
requested (of 0\farcs8, \S~\ref{sec:obsdatared}).

Fig.~\ref{fig:hst_vimos} helps assessing the accuracy of our HST to
VIMOS matching, and further shows the final map for the values of the
disc to total ratio in the VIMOS resolution elements.
The latter was obtained by simply applying the same rotation,
convolution and resampling steps (using the previously derived best
double-Gaussian PSF) to the best-fitting disc model image that was
derived by \citeauthor{Mor04}, and by then dividing the result by the
degraded HST image of NGC~4458.
Such a disc-to-total ratio map shows that in the central VIMOS spectra
we can expect a disc contribution nearly up to 5\%.

\subsection{Bulge Spectrum Analysis}
\label{subsec:analysisBulge}

To better constrain the age of the NSD in the central region where its
light contribution is the greatest, we ought to also have the best
possible model for the bulge stellar spectrum, which will nonetheless
dominate the central spectrum we are about to analyse.
For this we combined two bulge spectra extracted within two 3$\times$3
pixels apertures in opposite directions 2\farcs5 away from the centre
along the major axis of NGC~4458 (that is, horizontally in
Fig.~\ref{fig:hst_vimos}). Since we aim to reduce the impact of
stellar population gradients in the bulge, these are indeed the
closest regions to the centre where the contribution of the nuclear
disc is negligible, well below 1\%, the level at 2\arcsec, as can be
seen on the lower right panel of Fig.~\ref{fig:hst_vimos}. Were it not
for the peculiar vertically-elongated character of the VIMOS PSF, we
would have extracted our bulge aperture spectrum along the minor axis.

Following the extraction of the such a representative bulge spectrum,
we proceeded to match it in the best possible way using pPXF and the
entire MILES stellar library. Using all the 985 stellar spectra in the
MILES library allowed us to match also those spectral features that
are notably hard to reproduce when using stellar population synthesis
models owing to abundance patterns that can only be partially
accounted for in these templates \citep[e.g., the Mgb region; see
  also][]{Sar10}. In this fit we allowed for interstellar extinction
(adopting a \citealp{Cal00} reddening law) and for an additional
fourth-order additive polynomial correction of the stellar
continuum. The weights assigned to each MILES template during the pPXF
fit were then used to construct an optimal template for the bulge
stellar population.

\placefigfive
\placefigsix
\placefigseven
\placefigeight

\subsection{Nuclear Stellar Population Analysis}
\label{subsec:analysisNSD}

To estimate the age of the nuclear disc in NGC~4458, we extracted a
3$\times$1 pixels central aperture along the minor axis (that is,
vertically in Fig.~\ref{fig:hst_vimos}) where the disc-light
contribution (in the F555W filter wavelength region) amounts to $\sim
5\%$ and the $S/N$ per pixel reaches values of 120.
We then used pPXF to fit such a nuclear spectrum with the our bulge
template and the single-age stellar population models from both the
\citet{Mar11} and \citet{Vaz10} libraries (both based on the MILES
spectral atlas) in order to represent the disc, which we indeed assume
to have formed very quickly.
We constrained the age of the nuclear disc by considering, at a given
stellar metallicity, each one of the population models for the disc at
a time, combining them with our empirical bulge template in two
different ways. In the first approach (the free fit), we allow pPXF to
choose freely the relative weight of these two templates, whereas in
the second (the constrained fit) we accounted for the relative light
contribution that bulge and disc templates should contribute to the
nuclear spectrum, that is 95\% and 5\% for the bulge and disc,
respectively.
More specifically, since the disc-light contribution within our
central VIMOS aperture that we derived in \S~\ref{subsec:analysisHST}
refers to the light fraction in the same band-pass of the HST images
for NGC~4458, prior to our pPXF fit we integrated the flux of our
templates within the F555W passband (accounting for the fact that the
F555W filter response tails off to longer wavelength than the VIMOS
data), and then used such integrated fluxes when weighting them during
the constrained fit. During this preliminary step, we also reddened
each of the model templates and the bulge template by our best
measurement of the interstellar extinction toward the nucleus. This
reddening estimate was obtained by performing a pPXF fit to the
nuclear spectrum using the entire MILES stellar library as done
previously for the bulge aperture, a fit that will also set the
standard for the best possible model for our nuclear spectrum.

Using first the \citeauthor{Mar11} models,
Fig.~\ref{fig:first_results} shows the results of this exercise, where
the quality of the pPXF fit in the free and constrained cases are
compared as a function of stellar-population age for the disc
template, in this case, of Solar metallicity. In both instances, the
formal uncertainties in the flux density values of the nuclear spectra
where rescaled in order for our best possible fit, the one obtained
using the entire MILES library, to have $\chi^2$ value equal to the
number of degree of freedom of our fit (NDOF).
Both free and constrained approaches indicate that the nuclear disc
must indeed be very old, possibly as old as the bulge. The free pPXF
fit indeed always prefer to use ony our bulge template, which is
itself best fitted using the oldest and slightly metal-poor (half
Solar) of our single-age templates (consistent with the results of
both \citealp{Mor04} and \citealp{Kun10}), whereas during the
constrain approach only the oldest disk templates lead to similarly
good fits.
The behaviour of the free and constrained fits in
Fig.~\ref{fig:first_results} is more dramatic than that shown in our
initial simulations, and is due to the presence of an overabundance
pattern in alpha elements in our nuclear data, in particular in the
spectral region corresponding to the Mgb Lick index
\citep{Gon93}. This abundance pattern is partially accounted for by
our empirical bulge template, and the fit in these regions is only
made worse during the constrained fit where the quality of the fit
quickly deteriorates as we force the use of a progressively younger
disc template that contributes to $\sim 5\%$ of the light in the
nuclear spectrum \citep[as expected from the disc-bulge decomposition
  of][]{Mor04}.

In Fig.~\ref{fig:first_results} our rescaling of the formal
uncertainties on the flux density in our spectra is useful to show
that the our models for estimating the age of the disc cannot quite
match the quality of a fit that allows for any possible mix (even
unphysical ones) of stellar spectra entering our nuclear spectrum of
NGC~4458.
In fact, even when adopting such an empirical description for the
bulge, the use our empirical template, with or without the addition
single-age stellar population models for the disc, we obtain fits that
are $\ga$40\% worse than our best fit. This may highlight the presence
of a substantial central gradients in the properties of the bulge
populations that we cannot capture with our bulge template,
limitations in the spectral synthesis models that we adopt for the
disk, or that the disc formation was instead rather prolonged.
Given that our tools of trade are not optimal, in order to be as
conservative as possible in placing a lower limit on the age of the
disc based on $\Delta\chi^2$ statistics, we ought to further
artificially broaden our flux-density errors until our best
constrained pPXF fit becomes formally a good fit (i.e., until the
corresponding $\chi^2$ reaches down to NDOF).
In fact, for our final estimate of the disc age, we decided to further
restrict our analysis to the spectral regions around the age-sensitive
H$\delta$ and H$\beta$ stellar absorption features. This does not
bring much loss of information, since most of the difference between
our models occur in these spectral regions (from 90\% to 40\% of the
quadratic difference when comparing models including young and old
disc models or for old disc ages only, respectively; see also
Fig.~\ref{fig:model_diffs}), and has the advantage of making our age
estimates less sensitive to the way the polynomials adjust the
continuum shape of our models, on which we have little control during
the pPXF fit.
Finally, we considered single-age stellar population models for the
nuclear disc of half and twice Solar metallicity, in addition to the
Solar metallicity models used in our first attempt of
Fig.~\ref{fig:first_results}.

Fig.~\ref{fig:final_results} shows the run of the quality of our
constrained fit in the H$\delta$ and H$\beta$ spectral windows as a
function of the age of the disc population model that is combined with
our empirical bulge template, with different lines indicating the use
of single-age models of different stellar metallicity.
As anticipated above, all formal uncertainties on the flux density
values of our nuclear spectra have been rescaled until our best
constrained pPXF fit - in this case including a stellar disc of
super-Solar metallicity - lead to a $\chi^2$ value equal to NDOF
across the entire wavelength range.
The $\chi^2$ values plotted in Fig.~\ref{fig:final_results} correspond
then only to the portion of our spectra within 30~\AA\ of the
H$\delta$ and H$\beta$ absorption lines at 4340~\AA\ and 4861~\AA,
respectively.
Setting a 1-parameter $\Delta\chi^2=9$ bar above the $\chi^2$ value of
our best-fitting constrained model allows to finally place a $3\sigma$
lower limit of $\sim$5--6 Gyr on the stellar age of the nuclear disc
of NGC~4458.
Using the whole spectral range would have yielded a tighter lower
limit of $\sim$7--8 Gyr. We also note that possible nebular in-fill
contamination in the H$\delta$ and H$\beta$ spectral regions is
excluded as there is little evidence for [{\sc
    O$\,$iii}]$\lambda\lambda4859,5007$ emission in the central
regions of NGC~4458, either from our data (see Fig.~\ref{fig:censpec})
or in the SAURON integral-field data shown by \citet{Sar06}.
To conclude, we note that our result do not depend on the choice of
stellar population models. For instance,
Fig.~\ref{fig:final_results_M11} shows that using the \citet{Vaz10}
single-age models for representing the nuclear disk population lead to
very similar old age constraints on the disk age. This may have not
been the case if the nuclear disk had turned out to be substantially
youngers, since model prescription can vary in this case
\citep[e.g.][]{Mar05}.

\section{Discussion}
\label{sec:discuss}

By combining high-quality VIMOS integral-field spectroscopic
observations with constraints from HST images on the relative
contribution of the nuclear disc of NGC~4458 to the central surface
brightness of this galaxy we have been able to set a tight limit on
the stellar age of such NSD.
Our analysis indicates that its formation must have occurred at least
$\sim$5 -- 6 Gyr ago, which in turn suggests that NGC~4458 did not
experience any major merger event since that time.

Besides serving as a proof of concept for further measurements in
larger samples of NSD-hosting galaxies that could lead to a better
understanding of their assembly history, the finding of such an old
NSD in NGC~4458 already provides food for thoughts on the formation of
the specific class of early-type galaxies that display very little or
no bulk rotation.
Over the course of the SAURON survey \citep{deZ02} NGC~4458 was in
fact classified as one of those so-called slow-rotators \citep{Ems07},
which the ATLAS$^{\rm 3D}$ survey \citep{Cap11} firmly recognised as
forming only a minority, $\sim$14\%, of the entire early-type galaxy
population \citep{Ems11}.
More specifically, NGC~4458 falls in the kind of slow-rotators that
exhibit a central slowly-rotating core within a non-rotating main
stellar body (also known as a kinematically decoupled core, KDC).
As in the case of other galaxies in this class, the kinematic
transition to the rotating core (which in the case of NGC~4458 occurs
$\sim$5\arcsec\ from the centre) does not appear related to any
noticeable photometric or stellar-population feature, except that in
the case of NGC~4458 a NSD is further found well within it (at
1\arcsec\ scales).

The formation of slow-rotators is still an open issue for theoretical
models.
Indeed, whereas from a simple semi-analytical approach the present-day
relative fraction of fast and slowly-rotating early-type galaxies is
well reproduced by considering as fast rotators all objects that in
these models have at least 10\% of their total stellar mass in a disc
component \citep[thanks to a more prolonged gas accretion
  history,][]{Kho11}, numerical simulations for galaxy interactions
still have a hard time reproducing both the kinematic and photometric
properties of slow rotators.
Under certain conditions binary mergers between discs can lead to
remnants resembling slow-rotators with a KDC \citep{Jes09}, but
generally such simulated objects are much flatter than real
slow-rotators. Additional major merger encounters do not address such
a discrepancy, but instead destroy the central core and lead to an
overall larger angular momentum \citep{Boi11}.
In fact, it is generally difficult to decrease the stellar angular
momentum through major mergers since these encounters bring a great
deal of orbital angular momentum that must be conserved
\citep{Kho06b}.
For this reason, frequent minor mergers have been advocated as a more
efficient means for both removing the angular momentum of galaxies and
making them rounder \citep{Kho11}.
Yet, even though the negative impact of minor mergers on the angular
momentum and flattening has been observed in several numerical
simulations carried out in a cosmological context \citep{Naa14}, there
is still limited agreement as regards the intrinsic flattening of
simulated and real early-type galaxies (for the latter, see
\citealp{Wei14}, based on the ATLAS$^{\rm 3D}$ sample).

In this respect we note that NGC~4458, with its perfectly edge-on
nuclear disc that presumably sits in the equatorial plane, must
intrinsically be a nearly spherical galaxy given its apparent axis
ratio $b/a = 1-\epsilon = 0.88$ \citep[where the flattening $\epsilon$
  is from][]{Ems11}. Furthermore, NGC~4458 is special among
slow-rotators, in that it is the least massive object in this class.
Its dynamical mass is estimated at $10^{10} \rm M_{\odot}$
\citep{Cap13}, whereas most non-rotators and slow-rotators with a KDC
have mass values of $1.8\times 10^{11} \rm M_{\odot}$ \citep{Ems11}.
Both these characteristics make NGC~4458 particularly puzzling.
Indeed if the roundness of NGC~4458 could suggest that minor mergers
were particularly important in shaping it, its small mass would argue
against it since presumably only the most massive systems would have
seen many smaller galaxies coming their way during their history.
In addition, the odds of NGC~4458 interacting or merging with other
galaxies would have further decreased further since it entered the
Virgo cluster.
The mere presence of a NSD in NGC~4458 could represent an additional
argument against a late satellite bombardment \citep[although minor
  mergers may not always effect the central regions of a galaxy,
  see][]{Cal09,Cal11}, whereas the old age of the NSD is consistent
with the notion that NGC~4458 did not experience a major merger in a
long time and hence had most of its mass in place early-on in its
history.
Finally, NGC~4458 also hosts a KDC which, as suggested already by many
authors for this kind of structures
\citep[e.g.,][]{Bal90,Her91,DiM08,Boi11}, could have formed during a
gas-poor merger event.
The presence of a NSD embedded in such a KDC indicates either that the
merger event that lead to the formation of the KDC must have preceded
the formation of the disc, or alternatively that the nuclear disk
could have also formed during such merger thanks to the presence of
some gas material. In fact, our analysis also allows for a co-eval
formation for the stars encompassed by the nuclear and bulge aperture,
which indeed covers the KDC regions.

\subsection{Caveats}

In our analysis we have assumed that the NSD of NGC~4458 formed almost
instantaneously, although this does not necessarily have to be the
case. If the star-formation history of the NSD was indeed prolonged
this is likely to have been characterised by a number of starbursts
each lasting just a few Myr, comparable to the short dynamical
timescales of galactic nuclei and consistent with the finding that
central starbursts consume their gas reservoir very efficiently
\citep[e.g.,][and references therein]{Bou11}. This is similar to the
case of nuclear clusters \citep{Bok02,Bok04}, which display optical
spectra that are indeed well matched by a superposition of single-age
stellar-population models \citep{Sar05,Wal05,Ros06}.
In the case of nuclear clusters the presence of several stellar
sub-populations in nuclear clusters could be investigated only because
nuclear clusters are generally young systems, with average ages
typically less than a few Gyrs. For the NSD of NGC4458, however, our
analysis indicates that the bulk of the stars in the NSD are very old,
a result that is robust even when using single-age stellar population
models. Indeed, if a significant fraction of young (less than 1--2
Gyr) stars was present in the NSD we would have otherwise inferred a
biased luminosity-weighted younger age for it. Given the difficulties
in separating old stellar populations from each other, we consider
presently unfeasible even with our photometric constraints to
disentangle the presence of distinct but similarly old episodes of
star formation in the NSD of NGC4458.

\subsection{Future outlook}
\label{subsec:discussfuture}

The results of our stellar-population analysis for the central regions
of NGC4458 demonstrate the accuracy with which it is possible to
constrain the age of NSDs with integral-field data. Considering that
such structures are present in up to 20\% of early-type galaxies
\citep{Led10} and in few spirals \citep{Piz02}, a similar but
systematic investigation of NSDs in galaxies of different mass and
across different galactic environments would constitute a promising
avenue for constraining the assembly history of early-type galaxies.
In fact, \citeauthor{Led10} already provides the most extensive sample
of nearby NSDs on which such a follow-up survey could be based, with
the recent work of \citet{Cor16} already finding an example of a young
NSD among the \citeauthor{Led10} sample.

In this respect, the MUSE integral-field spectrograph \citep{Bac10},
recently mounted at the Yepun-UT4 of VLT, will be particularly suited
for such an NSD census, for several reasons.
MUSE is indeed much more efficient than VIMOS (with an overall
throughput reaching up to $\sim$40\%) and extends to a longer
wavelength range (from 4650~\AA\ to 9300~\AA), allowing a neater
separation of the nebular emission from the stellar continuum while
retaining a spectral resolution of R$\sim$3000 sufficient for a
detailed estimate of the stellar LOSVD.
Finally, and most importantly for the study of NSDs, MUSE will
eventually also work with adaptive optics and reach a spatial
resolving power comparable to that of HST, which will dramatically
boost the disc contribution to the nuclear spectra.
This will lead to even tighter constraints of the NSD age, in
particular if the stellar kinematics observed in the nuclear regions
will be also brought in as an additional constraint.

A better NSD-to-bulge contrast and the ability to constrain the
stellar LOSVD (in turn thanks to an excellent data quality and of a
respectable spectral resolution) should allow to fold into our
analysis a self-consistent dynamical model for the central kinematics
\citep[e.g. based on Jeans equations,][]{Mag99,Cap08}, which will be
also very sensitive to the age of the disc.
For instance, at a given disc-light contribution, choosing an older
stellar population for the disc will mean considering a more massive
nuclear disc that will imprint a larger rotation velocity to the disc
stars in the models, whereas picking a very young age will translate
into a disc dynamics almost entirely determined by the gravitational
potential of the bulge.

If our ability to constrain the stellar age of nearby NSDs seem set to
improve in the near future, steps will also have to be made on a more
theoretical side in order to better understand the implications of
such age estimates.
In particular it will be paramount to assess the extent to which NSDs
are fragile to merger episodes, so that the presence of a NSDs can be
firmly translated into a maximum mass ratio for any accretion event
that could have followed the formation of the disc.
At the same time, it will also be interesting to follow the disruption
of NSDs during more dramatic encounters, looking for instance for the
possible kinematic signature of the past presence of such structures.
Progress in this direction has already been made by Sarzi, Ledo \&
Dotti (2015) using a relatively large set of numerical simulations,
which show not only how NDSs emerge relatively unscathed from minor
mergers (e.g., for a 1 to 10 mass ratio or less) but also that a
central rotating structure could still be present at the end of more
important interactions that leave no photometric trace of the NSD.
Although encouraging, these results are still based on simple initial
conditions mimicking the final phases of a merger event and which
always lead the the central black hole of the satellite galaxy to sink
towards the centre, whereas this may not always be the case as we
already noted.
More comprehensive simulations are needed to fully understand both the
origin of NSDs \citep[][]{Por13,Col14} and their fragility against
minor mergers, possibly shedding also more light on the origin of
kinematically decoupled central structures and the possible link to
past central discs.

Finally, we note that the advantage of knowing \textit{a priori} the
stellar light contribution of a given stellar subpopulation to the
optical spectra of a galaxy may be used to constrain also the stellar
age of other kinds of galactic components beside NSDs, such as more
extended discs \citep[see, e.g.,][]{Joh12,Coc14,Joh14,Coc15}, nuclear
rings or other photometrically distinguishable structures.

\section*{Acknowledgements}

We are grateful to anonymous referee for carefully reading our
manuscript and for providing comments that helped improving this work.

\label{lastpage}
\end{document}